%% file: draft_10.tex
\documentclass[journal=ancac3,manuscript=article,layout=traditional]{achemso}

\usepackage{amsmath}
\usepackage{amsfonts}
\usepackage{amssymb}
\usepackage{graphicx}

\usepackage{xspace}
\usepackage{bm}
\usepackage{booktabs}

\newcommand{\vc}[1]{\textbf{#1}}

\usepackage{titlecaps}
\input{titlecase_exclude}

\renewcommand{\d}[0]{\ensuremath{\,\mathrm{d}}\xspace}

\newcommand{\onlinecite}[1]{\hspace{-1 ex} \nocite{#1}\citenum{#1}}

\newcommand{\visc}[0]{\ensuremath{\eta}\xspace}
\renewcommand{\u}[1]{\ensuremath{\,\mathrm{#1}}\xspace}
\newcommand{\pmma}[0]{\textsc{pmma}\xspace}
\newcommand{\dcm}[0]{\textsc{DCM}\xspace}
\newcommand{\via}[0]{\ensuremath{\mathit{via}}\xspace}

\title{Accurate Location and Manipulation of Nano-Scaled Objects Buried under Spin-Coated Films}

\author{Colin Rawlings}
\author{Heiko Wolf}
\affiliation[IBM Research - Zurich]{IBM Research - Zurich, Rueschlikon 8803, Switzerland}
\author{James Hedrick}
\author{Dan Coady}
\affiliation[IBM Research - Almaden]{IBM Research - Almaden, San Jose, CA 95120 USA.}
\author{Urs Duerig}
\author{Armin Knoll}
\email{ark@zurich.ibm.com}
\affiliation[IBM Research - Zurich]{IBM Research - Zurich, Rueschlikon 8803, Switzerland}

\date{\today}

\begin{document}

\begin{abstract}
Detection and precise localization of nano-scale structures buried beneath spin coated films are highly valuable additions to nano-fabrication technology.
In principle, the topography of the final film contains information about the location of the buried features. However, it is generally believed that the relation is masked by flow effects, which lead to an upstream shift of the dry film's topography and render precise localization impossible.
Here we demonstrate, theoretically and experimentally, that the flow-shift paradigm does not apply at the sub-micron scale. Specifically, we show that the resist topography is accurately obtained from a convolution operation with a symmetric Gaussian Kernel whose parameters solely depend on the resist characteristics. We exploit this finding for a 3\u{nm} precise overlay fabrication of metal contacts to an InAs nanowire with a diameter of 27\u{nm} using thermal scanning probe lithography. 
%
%
\end{abstract}

\maketitle


The identification of buried features is of interest in a wide range of fields, including biology \cite{Tetard2008}, polymer physics \cite{Spitzner2011} and semiconductor fabrication \cite{Shekhawat2005}.  A number of techniques exist for detecting micron-scaled features, for example sonography and confocal microscopy.  However, nondestructive options at the nanometer scale are more limited. Transmission Electron Microscopy has been shown to be capable of imaging carbon nanotubes in cells \cite{Porter2007}.  Methods based on the ubiquitous atomic force microscope offer the potential for nondestructive imaging at nanometer resolution without the need for careful sample preparation.  An early AFM-based technique capable of detecting buried features employed a high-frequency acoustic wave \cite{Cuberes2000} and a tip detection scheme.  Recently, it has been shown that the amplitude modulation operating mode of AFM can be used to detect buried features \cite{Spitzner2011,Ebeling2013} beneath soft films.

\begin{figure*}[t]
  \centering
  \includegraphics[width=0.9\textwidth]{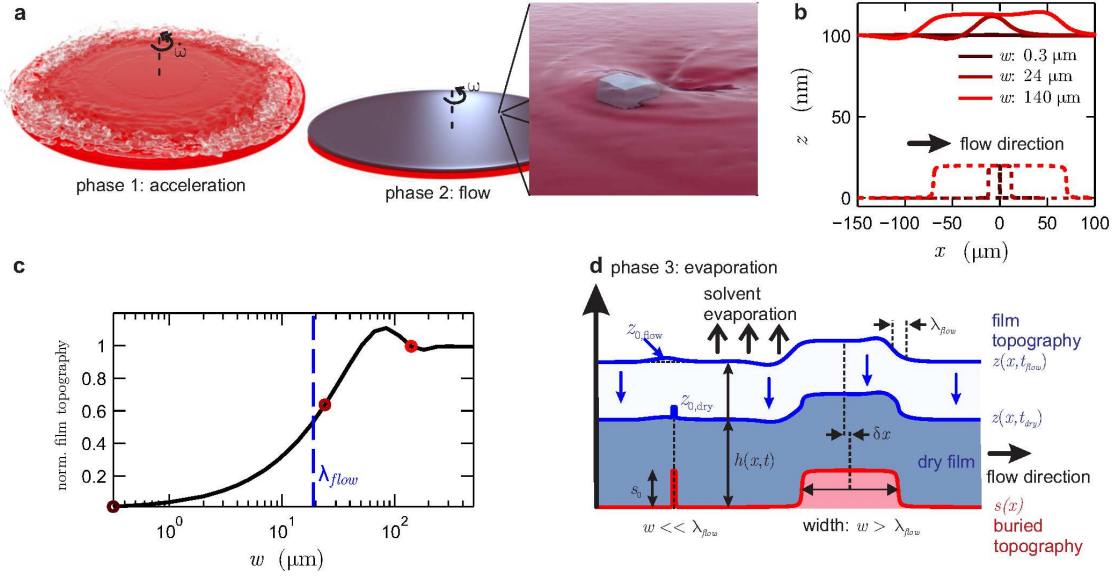}
  \caption{Diagram showing the formation of the film topography during the three phases of spin coating of a polymer solution over a wafer with existing topography.  (a) Schematic representation of the first two phases of spin coating a polymer film: the chaotic acceleration phase as the disc begins to rotate and the steady-flow phase.  The perturbation of the moving liquid film due to a wide sub-surface feature is shown in the enlargement.  (b) Calculated free surface $z$ of the polymer solution  (solid line) when flowing over wires (dashed lines) of different widths $w$ orientated perpendicular to the flow (radial) direction (see text and methods section for details).
  (c)  The dependence of the film topography's amplitude at $x=0$ on the wire width $w$.  The amplitude has been normalised with respect to the height of the sub-surface feature.  The circles correspond to the cross sections of (b).  The length scale for the  flow, $\lambda_{flow}$ (see equation (\ref{eq:lflow})), is shown by the blue dashed line.  (d)  Schematic contrasting the formation of film topography during the flow $z(x,t_{flow})$ and evaporation $z(x,t_{dry})$ phases over  narrow ($w < \lambda_{flow}$) and wide ($w>\lambda_{flow}$) buried features.   For narrow features, the topography emerges because of the shrinkage of the film's thickness $h$ and the topography is in registry with the underlying feature ($\delta x = 0$). }
  \label{fig:cartoon}
\end{figure*}

%
The detection of objects buried beneath spin-coated films is of particular interest given the widespread use of spin coating to prepare resist films with precisely controlled thickness. As a consequence, spin coating resides at the heart of micro- and nanofabrication and is part of many of the standard processes, in particular those involving lithography, such as pattern transfer, metal lift-off and local ion implantation. It has been shown that substrate topography leads to a disturbance of the surface of the spin-coated film \cite{White1985,Peurrung1991,Stillwagon1987,Stillwagon1988,Stillwagon1990,Gu1995,Kucherenko2000,Wu1999,Wang1995,Sukanek1989}. The effect impairs the flatness or planarity of the final film and has long been of concern to the semiconductor industry.  Previous work has shown that for large ($\geq \mathcal{O}(10\u{\mu m})$) features, this relationship  strongly depends on the local direction and rate of fluid flow during spin coating \cite{Peurrung1993,Hayes2000} (see figure \ref{fig:cartoon}).

However, this residual topography in principal offers a straightforward means of locating existing structures buried beneath the spin-cast film using surface-sensitive techniques, such as scanning probe microscopy (SPM).  Provided nanometer precision in the detection of buried feature position is possible, this approach can assist in the development of a quantitative understanding of other sub-surface techniques \cite{Verbiest2012,Bosse2014} on model substrates.  Moreover, precise location of features buried beneath spin-coated films would provide a direct means for addressing the critical challenge of overlay accuracy for device fabrication.  Scanning probe lithography (SPL) tools \cite{Garcia2014} can readily measure nanometer-scale topography as well as transfer sub-20\u{nm} features into silicon, \cite{Wolf2014} and thus provide a unique method for nanometer-precise device fabrication.

\clearpage

In this paper, we demonstrate that for sub-micron sized features the relationship between surface and sub-surface topography becomes independent of the flow direction.
We show analytically that for the limiting case of small sub-surface topography the surface topography may be obtained by convolving this sub-surface topography with a kernel that is well approximated by a Gaussian.
The model is then validated for the widely used resist Polymethylmethacrylate (\pmma) for in-plane length-scales spanning 10-1000\,nm. We show that the residual topography may be predicted after calibration of the resist-solvent system and further that the topography may be enhanced by an additional solvent annealing and quenching step. Finally, we demonstrate the utility of this approach for addressing nano-scale devices by scanning probe lithography.

\section{Results and Discussion}

\subsection{Phases of Spin Coating}
\label{sec:phases-spin-coating}

The spin-coating process may be divided into three distinct phases \cite{Lawrence1988}, see Figure \ref{fig:cartoon}.  The first phase  involves the rapid angular acceleration of the dispensed spin-coating solution as the substrate begins to rotate \cite{Rehg1988}.  Under the influence of the inertial forces, the fluid spreads on the surface. When the liquid layer has become sufficiently thin, \cite{Matar2005} the so-called ``flow phase'' begins, in which the fluid moves radially at a low Reynolds number under the influence of the inertial force \cite{Lawrence1988}.
The strong dependence of the flow rate on the film thickness $h$ causes the latter  to progress towards a conformal coating of the wafer \cite{Emslie1958}.  If topography is present on the surface a competition arises between the surface tension, which seeks to level the film, and the inertial term, which seeks to maintain a uniform film thickness \cite{Hwang1989}, see Figure \ref{fig:cartoon}b.
The final ``evaporation phase'', which was identified by Meyerhofer \cite{Meyerhofer1978}, follows the flow phase.  In the evaporation phase, flow is negligible and film thinning is dominated by solvent evaporation, see Figure \ref{fig:cartoon}d.  %

The significant difference between the in-plane and out-of-plane length scales for the film encountered in spin coating allows the reduction of the Navier Stokes and continuity equation to the so-called Lubrication approximation \cite{Howison2005} for the film thickness ($h$):
\begin{equation}
\label{eq:lub}
\partial_t h = -\frac{\gamma}{3 \visc} \nabla.\left( h^3 \nabla \left( \nabla^2 s +\nabla^2 h \right) \right) + j,
\end{equation}
where $\gamma$ is the surface tension, \visc is the fluid viscosity, $s(x,y)$ (see figure \ref{fig:cartoon}a) is the buried topography and $j$ is the volume of evaporated solvent per unit area \cite{Wang1995}.
For the case of spin coating, a body force term $-\rho \omega^2 r_0 \vc e_r$ is added to account for the rotation of the film's co-ordinate system.
The lubrication approximation has been found to be accurate to better than 10\% peak error even for flows over sharp steps whose height is on the order of the film thickness \cite{Mazouchi2001,Gaskell2004,Veremieiev2010}.  %
Previous workers have solved the coupled system of equations numerically for flat surfaces  \cite{Meyerhofer1978,Lawrence1988,Bornside1989,Matar2005,Munch2011} and surfaces exhibiting radial symmetry but containing $\mathcal{O}$(10\u{\mu m})-sized features \cite{Gu1995,Kucherenko2000}.

Useful insight into the relevant in-plane length scale for the flow phase ($\lambda_{flow}$) may be obtained from the approximate analytical solutions of refs. \onlinecite{Stillwagon1990} and \onlinecite{Wu1999} for the flow phase.  The authors show that the flow over a radially symmetric step located at $x=0$ may be constructed from a set of three linearly independent terms which decay as $\exp(-\alpha |x| / \lambda_{flow})$.  $\alpha$ is a constant of order unity and $\lambda_{flow}$ is given by 
\begin{equation}
\label{eq:lflow}
\lambda_{flow} = \sqrt[3]{\frac{\gamma h}{\rho \omega^2 r_0}},
\end{equation}
where $\rho$ is the fluid density, $\omega$ is the sample's angular velocity, and $r_0$ is the radial position of the buried feature.  Substituting reasonable values of $\gamma=30$\,Nm$^{-1}$, $h=100$\,nm,  $\omega=2000$\,rpm, $\rho=1000$\,kgm$^{-3}$ and $r_0=10$\,mm yields $\lambda_{flow}=18$\,$\mu$m.  The results of a numerical calculation of the flow over a radially symmetric feature with a rectangular cross section are shown in  figures \ref{fig:cartoon}b and \ref{fig:cartoon}c (see methods section for details).  Features whose width $w$ exceeds $\lambda_{flow}$ result in a change in height of the fluid's upper surface that is equal to the height of the buried feature.   Conversely, if $w \ll \lambda_{flow}$ the free surface remains almost flat above the buried feature.
%


\subsection{Final Topography: Small Structures ($w \ll  \lambda_{flow}$)}

For features whose width is less than $\simeq 1\mu$m (see figure \ref{fig:cartoon}c), the topography present in the film surface as the evaporation phase begins is a few percent of the sub-surface topography.  As such, its contribution to the surface topography of the dry film is dominated by the additional topography emerging during the evaporation phase.  Previous workers \cite{Peurrung1991,Stillwagon1990} have made the  assumption that the topography formed during the evaporation phase may be calculated by assuming a uniform shrinkage of the film present as the evaporation phase starts.  Here we refine this assumption by observing that it predicts the transfer of sharp step-like features from the sub-surface topography to the film topography.  The development of such a sharply curved surface would be opposed by the surface tension, which would cause the film to flow, see Figure \ref{fig:theory}.  The competition between evaporation-driven topography formation and capillary flattening proceeds until the polymer matrix becomes glassy and solvent evaporation ceases.

\subsection{Analytical Solution}

We make several approximations to allow for an analytical treatment of the evaporation phase.  Firstly, the body-force term is assumed to be negligible during this phase.  Secondly, we retain the assumption used for large structures that the loss of solvent per unit area is proportional to the film thickness at the end of the flow phase.  Thus the evaporation term in equation (\ref{eq:lub}) becomes
\begin{equation}
\label{eq:j_approx}
j(x,y,t) = -q(t) (h_f-s(x,y)),
\end{equation}
\noindent
where $q(t)$ is the evaporation rate per unit initial film thickness during the evaporation phase, $t_f$ is the time at which the flow phase ends and $h_f$ is the film thickness far from any topographic features at $t=t_f$.

\begin{figure}[h]
  \centering
 \includegraphics[width=0.5\textwidth]{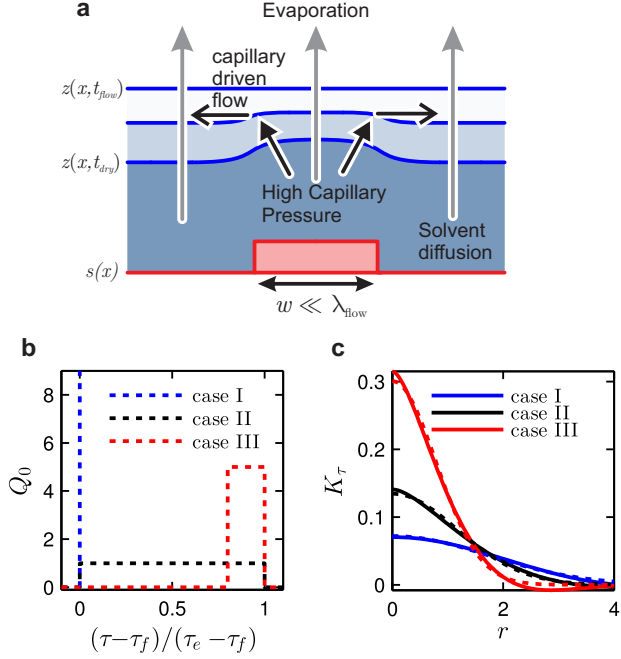}
  \caption{Model and analytical solution for the formation of film topography over small buried features during spin coating.
(a) Diagram showing the temporal evolution of the free surface (blue lines) due to the competing effects of volume loss due to solvent evaporation and film flow due to capillary forces.  (b) Three distinct profiles for the unknown function $Q_0$ in equation  (\ref{eq:Kt_f_Q0}) (n.b. Case I: $Q_0 = \delta(\tau-\tau_f)$).  (c) Form for the kernel function (solid line) $K_{\tau}$ for each of the cases of $Q_0$ shown in (b), the least squares fit of a Gaussian to each curve is also shown (dotted line).}
  \label{fig:theory}
\end{figure}

Equation (\ref{eq:lub}) can now be solved for the approximation of equation (\ref{eq:j_approx}) and for $s\,<<\,h$
 %
\via an asymptotic expansion.  The rapidly varying viscosity is treated as a known function of time $\visc(t)$ at this stage.  This yields the pair of equations for the first two terms in the expansion where $h=h_0(t) + h_1(x,y,t)$ and $z_1 = h_1 + s$:
\begin{eqnarray}
\partial_t h_0 = -q(t) h_f \label{eq:assym_exp_h0} \\
\partial_t h_1 = \partial_t z_1 = -\frac{\gamma h^3_0(t)}{3 \visc(t)} \nabla^4 z_1 + q(t) s \label{eq:assym_exp_q}
\end{eqnarray}
Given $q(t)$, equation (\ref{eq:assym_exp_h0}) may be integrated to obtain the uniform thinning of the film due to evaporation.  Equation (\ref{eq:assym_exp_q}) can also be readily integrated by first defining a re-scaled time parameter $\tau$  to accommodate the changing mobility of the film with $h_0(t)$ and $\visc(t)$:\footnote{The variation of the surface tension with solvent concentration can also be handled by conceptually making $\gamma(t)$.  This does not account for the $\mathcal{O}(s)$ in-plane variation of the concentration required to achieve the faster evaporation rate.}
\begin{equation}
\label{eq:2}
\tau(t): \quad\frac{d \tau}{dt} = \frac{\gamma h^3_0(t)}{3 \visc(t)}, \quad \tau_f = \tau(t_f) .
\end{equation}
Note that the viscosity eventually diverges so that $\tau$ goes to some finite value $\tau_e$ as $t\rightarrow \infty$ .
Substituting this change of variable into equation (\ref{eq:assym_exp_q}) yields
\begin{equation}
\label{eq:1}
\partial_{\tau} z_1 = -\nabla^4 z_1 + Q_0(\tau) s(x,y), \quad Q_0 = \frac{3 \visc}{\gamma h^3_0} q.
\end{equation}
The variable $Q_0$ describes the evaporation rate scaled by the film's ability to flow in the face of the emerging topography.  We are interested in the profile of the dry film $z_e$, which is given by
\begin{align}
\label{eq:3}
z_e(x,y) &= z_1(x,y,\tau_e) \\
 &= s(x,y) \ast K_{\tau} \\
K_\tau &= \int_{\tau_f}^{\tau_e} Q_0(\tau) K(x,y,\tau_e-\tau) \d \tau,
\label{eq:Kt_f_Q0}
\end{align}
where $\ast$ denotes convolution and $K$ is the Green's function for equation (\ref{eq:1}):
\begin{equation}
\label{eq:4}
K(x,y,\tau) =
\begin{cases}
 \frac1{2 \pi \sqrt{\tau}} \int_0^{\infty} u e^{-u^4} J_0 \left( \frac{u \sqrt{x^2+y^2}}{\sqrt[4]{\tau}}\right)  \d u  \\
0, \quad \tau < 0
\end{cases},
\end{equation}
where $J_0$ is a Bessel's function of the first kind.

The usefulness of preceeding analysis is derived from the insensitivity of the shape of $K_\tau$ to the form of $Q_0(\tau)$.  Figure \ref{fig:theory}b shows three quite different profiles for $Q_0$.  Figure \ref{fig:theory}c shows the corresponding values for $K_\tau$ for each case.  Each of these curves is well approximated by a Gaussian, with the exception of the evanescent small amplitude oscillations at large $r$.
  Thus $K_\tau$ may be described in terms of just two parameters, the Gaussian's width $\sigma$ and its amplitude $R$:
\begin{equation}
\label{eq:5}
K_{\tau} \simeq \frac{R}{\sigma^22\pi} e^{-(x^2+y^2)/(2\sigma^2)}
\end{equation}

Case III of figure \ref{fig:theory}b is likely closest to the physical $Q_0(\tau)$, which corresponds to a formation of topography towards the end of the evaporation phase when the film is relatively less able to level itself.
Initially the rate of solvent loss is controlled by evaporation from the film interface \cite{Lawrence1988,Gu1995}.  As such, $q$ in equation (\ref{eq:1}) will fall linearly with the solvent concentration $\phi$ \cite{Sparrow1960}.  
As $\phi$ falls, $\eta $ will increase as a power law \cite{Flack1984,Meyerhofer1978,Bornside1989,Kucherenko2000} in (1-$\phi$) with  an exponent $x$ of $2.3 \leq x \leq 4$.  It may also be readily shown that during the evaporation phase $h_0 \propto (1-\phi)^{-1}$.
At some value of $\phi$, or equivalently $\tau$, $\eta/h_0^3$ will, if plotted on a linear scale, rise rapidly.  This corresponds to the left-hand edge of the pulse shown in figure \ref{fig:theory}b.  The value of $Q_0$ will be limited by the falling $q$.  Its rate of reduction is accelerated by the declining diffusivity of the solvent in the solution.  Initially this diffusivity is rather insensitive to the falling concentration; \cite{Lawrence1988} however below a critical concentration, the diffusivity drops rapidly.  For solutions of Polystyrene and Toluene, this critical concentration was identified as $\simeq$20\% \cite{Gu1995}.  At this concentration it may be anticipated that the polymer matrix has begun to form and that $\tau$ has reached $\tau_e$, signaling the end of the evaporation phase.

In this section, we have derived a simple description of the relationship between the buried and the dry film topography after spin coating.
In contrast to solutions obtained on large structures, where the flow phase dominates topography formation, the film topography is predicted to be vertically aligned with that of the substrate.
In a complete calculation, the dependence of the solvent diffusivity and the solution viscosity must first be measured as a function of solvent concentration.  In contrast, here the model is parameterised by a measurement of $\sigma$ and $R$ for the film thickness of interest.


\subsection{Model Validation}

\begin{figure*}[t]
  \centering
   \includegraphics[width=\textwidth]{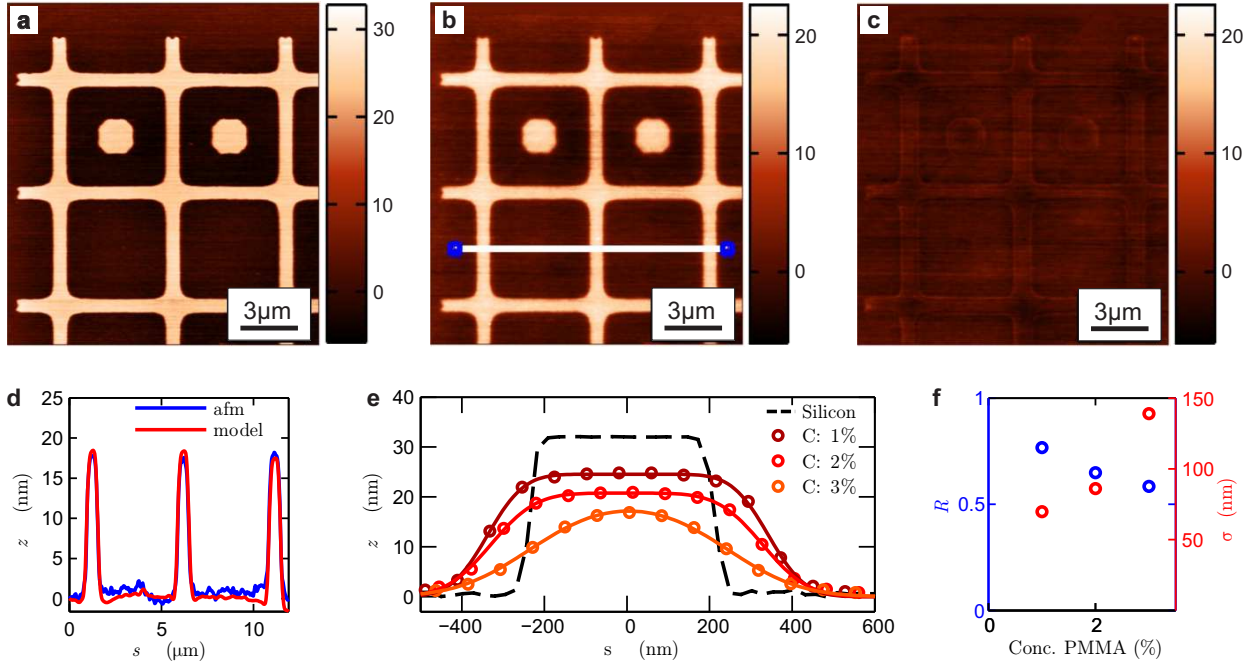}
  \caption{Experimental determination of $\sigma$ and $R$ which are required to parameterise the approximate form for $K_{\tau}$ [eq. (\ref{eq:5})] (a)  AFM image of topography etched into a silicon surface.  (b)  AFM  topography image following the spin coating of an 86nm \pmma layer.  The flow (radial) direction during the spin coating was from left to right.  (c) Difference between the measured topography of panel (b) and the topography predicted by the convolution of the topography in (a) with the fitted Gaussian Kernel.  The fit parameters [equation (\ref{eq:5})] were obtained as R = 0.69, $\sigma$ = 80nm.  (d) Comparison between the measured and predicted topography along the cross-section in (b).   (e)  AFM measurement (circles) of film topographies above a wire of rectangular cross section (black dashed line) following spin coating of \pmma solutions with concentrations from 1\% to 3\%.  The parameters in equation (\ref{eq:5}) which gave the best fit (solid line) to the experimental data are shown in (f). }
  \label{fig:expt1}
\end{figure*}

\begin{figure*}[t]
  \centering
 \includegraphics[width=\textwidth]{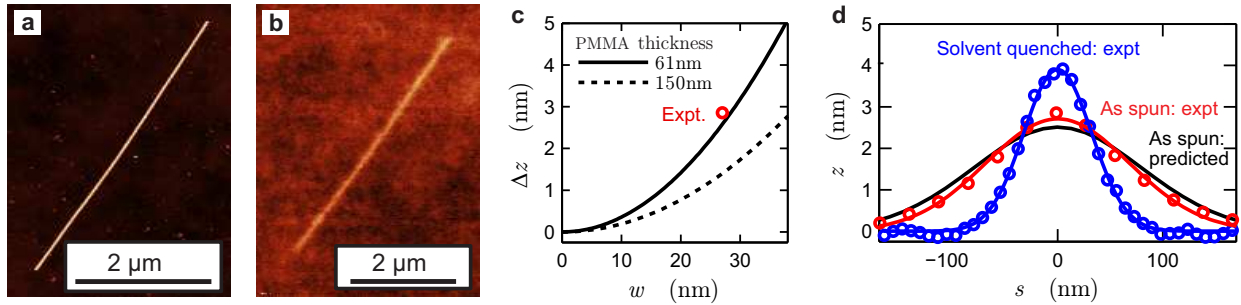}
  \caption{Prediction of residual topography when spin coating a 61\,nm \pmma film from an anisole solution over a 27nm diameter InAs nanowire.  (a) AFM image of an InAs nanowire on a silicon substrate, the diameter of the nanowire was obtained from the topography as 27nm.  (b) AFM measurement of film topography following spin coating of a 61nm \pmma layer over the nanowire of (a), the topography has an amplitude of 2.9nm.  (c)  Predicted topography amplitude as a function of nanowire diameter and film thickness.  The experimentally measured amplitude is shown by the red circle.   (d)  Nanowire cross-sections.  The measured cross-section of the nanowire is shown by the red circles.  The predicted topography using the parameters in figure \ref{fig:expt1}f is shown by the black line.  The blue circles show the cross section of the wire measured using an AFM after exposure of the film to a saturated dichloromethane (\dcm) atmosphere.  The solid red and blue lines show the least squares fit of a Gaussian to the experimental data.
}
  \label{fig:nwire}
\end{figure*}

 Figure \ref{fig:expt1}a and \ref{fig:expt1}b show AFM images of a patterned silicon surface before ($s$) and after ($z$) the addition of an 86\u{nm}-thick layer of \pmma \via spin coating.  The two images have been aligned by scratching the sample just below the portion of the AFM image shown in figures \ref{fig:expt1}a and \ref{fig:expt1}b (see figure S4).

From figures \ref{fig:expt1}c and \ref{fig:expt1}d, it can be seen that there is good agreement between the model's prediction and experiment for the obtained fit parameters of [see equation (\ref{eq:5})] $R$ = 0.69 and $\sigma=80\u{nm}$.  There is no observable translational offset between the surface and buried features.  
There is also no significant error irrespective of the wire orientation or near the ends of the wires, suggesting that the model is not solely able to predict the topography over wire-like features. Similar experiments have been performed with the commercial resist HM8006 manufactured by JSR Micro.  These experiments yielded results consistent with those obtained for PMMA (see figure S3).

Similar results were also obtained for solutions with concentrations varying from 1\% to 3\% spin coated over the wire structures shown in figure \ref{fig:expt1}a.  We measured the topography and used a least squares fit to identify the values of $\sigma$ and $R$ for each concentration, see figure \ref{fig:expt1}e and f.
 $\sigma$, which is representative of the length scale for the flow during the evaporation phase, increases with increasing film thickness.  This makes conceptual sense, given the strong ($h^{3}$) dependence of the flow rate on the film thickness which increases with concentration.  We likely do not observe a cubic dependence because the strong sensitivity of the viscosity on the concentration partially cancels the $h^3$ term.  The dependence of $R$ on the concentration also makes intuitive sense.  More concentrated solutions, which lead to thicker films, result in greater planarisation of the surface.

Using interpolation of the data presented in figure \ref{fig:expt1}f, the Kernel function for a film spun from a \pmma solution of arbitrary initial concentration may be calculated.  Knowledge of this Kernel in turn allows prediction of the dry-film topography for an arbitrary sub-surface topography, provided $w \ll \lambda_{flow}$.  To test the predictive power of our model, we dispersed InAs nanowires onto a silicon substrate.  One such wire having a diameter of 27\u{nm} and a length of roughly 6\u{\mu m} is shown in figure \ref{fig:nwire}a.  We spin coated a 60\u{nm} \pmma film onto the surface, which corresponded to a \pmma concentration of 1.5\%.  The surface topography of the wire after the spin-coating process is shown in figure \ref{fig:nwire}b.  The measured cross section of the wire is indicated by the red circles in figure \ref{fig:nwire}d.  Figure \ref{fig:nwire}c shows the predicted topography amplitude over the nanowire for two film thicknesses and a range of nanowire diameters.  Linear interpolation was used in the calculation, and the nanowire cross section was approximated as a square of side $w$.  The experimental result, corresponding to $w=27\u{nm}$, is also shown in figure \ref{fig:nwire}c.  The predicted cross section is shown as the black line in figure \ref{fig:nwire}d.

 There is good agreement between the predicted and measured cross sections.  The error in the peak height is just 10\%.   This is particularly encouraging because the Kernel was parameterised using measurements on a structure that was more than an order of magnitude larger than the nanowire.  The red line in figure \ref{fig:nwire}e shows the least squares fit of a Gaussian to this experimental data.  
 The high fit quality may be anticipated as the nanowire is much narrower than $K_{\tau}$.  As such, the topographic feature behaves like the Dirac distribution, so that the measured cross-section is approximately $\int_y K_{\tau} \d y$, which is itself a Gaussian function.
The result provides direct evidence that the form derived for $K_{\tau}$ is appropriate.  The value of $\sigma$ interpolated from the calibration data was 80\u{nm}, which is in good agreement with the fitted value of 68\u{nm}.  A similar accuracy could be achieved when treating the inverse problem of re-constructing the buried topography from the measured film topography\footnote{Of course, the usual limitations would apply to the recovery of spatial frequencies above $1/\sigma$.}.

We note that so-called tip convolution effects \cite{Markiewicz1994}, which are a well known source of artifacts in AFM images, will not affect the results discussed here.  Our homemade tSPL probes have a typical radius of 5\u{nm} (see figure S10a) which is significantly less than the measured value for $\sigma$.  Thus, as shown in the supporting information, the effect of tip convolution on our measured film topographies (see figures S9 and S10) is negligible.

Our spin-coating model predicts that  
the width of $K_{\tau}$ should be reduced if the evaporation process is sped up.  To investigate this experimentally, we swelled the film shown in figure \ref{fig:nwire}b in saturated dichloromethane (\dcm) vapour for several minutes.
The cross section of the nanowire was then re-measured using AFM.  The results of this measurement are shown by the blue circles in figure \ref{fig:nwire}d.  
Assuming that in the swollen state the capillary forces are able to flatten the surface of the film, we expect the same physics to control the topography formation as in the case of spin coating. 
Consequently, the cross section of the film topography above the wire (blue line, figure \ref{fig:nwire}d) is once again  well fitted by a Gaussian.
The fitted value of  $\sigma$ is smaller by a factor of 0.47, which is consistent with the higher vapor pressure of DCM (360 kPa) as compared to anisol (46 kPa).
This experimental result is of significant technological value.  Use of this volatile solvent has enhanced both the amplitude and the sharpness of the pattern, allowing it to be more accurately detected in the AFM.  Exposure after spinning avoids the difficulties associated with spinning from a highly volatile solvent that can lead to films exhibiting high surface roughness  \cite{Lai1979,Spangler1990}.

\begin{figure*}[t]
  \centering
\includegraphics[width=\textwidth]{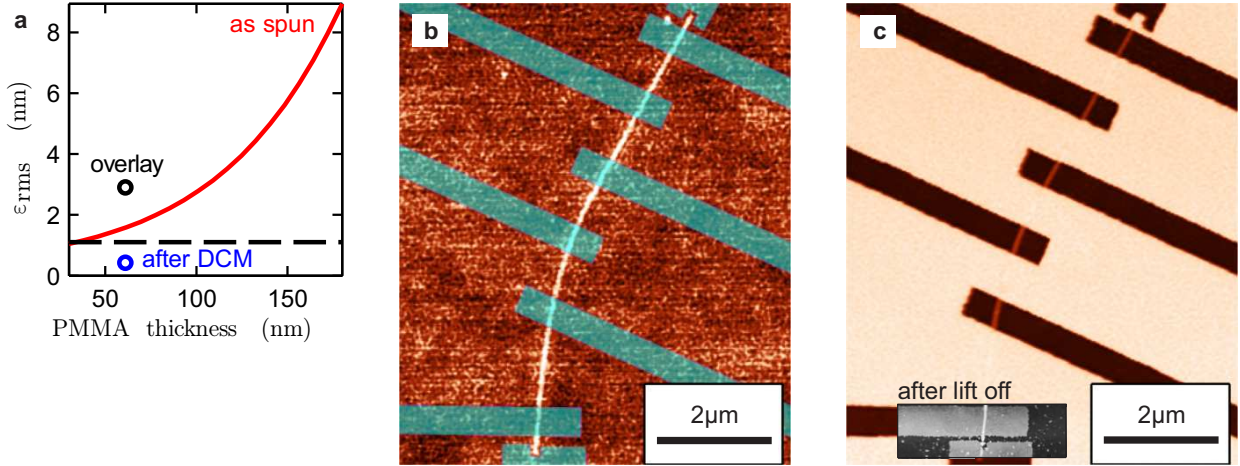}
  \caption{Demonstration of using tSPL to write contact pads in registry with a nanowire.  To enable pattern transfer the nanowire was placed beneath a pattern transfer stack \cite{Cheong2013} of [\pmma (61\,nm) $|$ SiOx (4\,nm) $|$ PPA (20\,nm)] (a) Predicted RMS correlation error $\varepsilon_{rms}$ when locating a nanowire of length 8$\mu$m and diameter 27\,nm which is buried under a \pmma layer having the characteristics shown in figure \ref{fig:expt1}f.  The calculation was performed using the approach detailed in ref.  \citenum{Rawlings2014}.  (b)  Topography image of the nanowire underneath the patterning stack.  The topography due to the wire has an amplitude of 3\,nm.  The design for the contact pads has been overlaid in blue. (c) AFM topography image following writing of the pattern using tSPL and subsequent pattern transfer into the \pmma \via a series of reactive ion etches.  The inset (gray colormap) shows a portion of the contact pads following lift off.}
  \label{fig:overlay}
\end{figure*}

The theory outlined so far provides a simple means of predicting how the topography over a buried feature will change as the thickness of the spin coated layer increases.  As the layer becomes thicker, the signal will become weaker and will eventually become lost in the detection noise.  The detection noise originates from the electronic noise of the instrument and the surface roughness of the spin-coated polymer \cite{Knoll2013}.  The problem of predicting the error in feature position due to noise has been considered previously, and details are given in ref. \onlinecite{Rawlings2014}.

\subsection{Application}

We close by applying our markerless overlay approach to the fabrication of electrodes used to contact a nanowire.   The transfer of a pattern from the resist layer  to the substrate provides a direct experimental means of confirming that the resist topography is not shifted with respect to buried features.

Figure \ref{fig:overlay}a shows the expected error when detecting the transverse position of the nanowire as a function of the film thickness \cite{Rawlings2014}.  A \pmma film thickness of 60\,nm was selected to balance the requirements of overlay accuracy and succesful lift-off of the evaporated metal film.  After the spin-coating process, the nanowire was exposed to the \dcm atmosphere, which resulted in an improvement in the amplitude and sharpness of the pattern.  These improvements reduced the expected position detection error when locating the position of the wire to less than 1\u{nm} (see figure \ref{fig:overlay}a).

Figure \ref{fig:overlay}b shows an AFM image of the nanowire taken by the tSPL tool prior to patterning.   The intended position of the contact pads with respect to the wire are shown by the transparent blue overlay.  Figure \ref{fig:overlay}c shows the device topography following tSPL patterning and etching.  The nanowire is visible at the bottom of each etched trench.  Figure \ref{fig:overlay}c has been aligned with the contact pad design shown in blue in figure \ref{fig:overlay}b using the cross-correlation based approach discussed in ref. \onlinecite{Rawlings2014}.  It can be seen qualitatively that the portions of the nanowire which are visible at the bottom of the etched trenches are aligned with the residual surface topography.    
Cross-correlation was used to locate the position of the exposed parts of the nanowire shown in figure \ref{fig:overlay}c with respect to the topography shown in figure \ref{fig:overlay}b (see figures S6-S8).  The overlay error perpendicular to the wire's long axis was measured as 2.9\u{nm}, which compares favourably with the wire's diameter of 27\u{nm}.  This value is somewhat larger than the expected correlation error.  This is likely due to three factors.  The first is the error in the tSPL instrument which has been measured as 1.1\u{nm} \cite{Rawlings2014}.  The second is that the sputtering of the SiOx layer damaged the \pmma surface, leading to a higher surface roughness\footnote{For subsequent devices, the SiOx was deposited using evaporation, which did not lead to an increase in surface roughness.}.  Finally, the reactive ion etch has introduced some edge roughness.  While this overlay error may not achieve the limit expected for the tool, it does provide further evidence that the topography of the dry film is precisely aligned with the sub-surface topograpy.  %

\section{Conclusion}
In this paper, we have shown theoretically and experimentally that spin coating across structures of feature widths $w \ll \lambda_{flow} \simeq 20 \u{\mu m}$ results in a residual topography that is vertically aligned to the buried features. This alignment occurs because at this length scale, the flow phase of the spin-coating process does contribute significantly to the topography of the dry film. Even for nanoscaled buried features the perturbation of the dry film topography may be readily measured using an AFM. The lateral position of buried features can therefore be determined using cross correlation with sub-nanometer accuracy. Good prediction of the final topography is possible by simply convolving the buried topography with a Gaussian Kernel. The Kernel may be parameterised through measurements on arbitrary features for which $w \ll \lambda_{flow}$. Using insight gained from the model, we have proposed a simple post-spin processing step whereby the dry film is exposed to a highly volatile solvent. In our case, this process reduced the detection error for the buried feature by a factor of three to a value of less than 1\u{nm}. Finally, we demonstrated the technological relevance of this work by performing markerless lithographic overlay onto a nanowire. We achieve an overlay error below 3\u{nm} or roughly one tenth of the nanowire diameter. This opens up a number of exciting possibilities in the experimental investigation of nanostructures.

\section{Methods}

\subsection{Calculation of fluid flows over wires}
\label{sec:calc-fluid-flows}
The profiles of figures \ref{fig:cartoon}b and \ref{fig:cartoon}c were obtained by solving equation (\ref{eq:lub}) for $j=0$.  It is shown in ref. \onlinecite{Stillwagon1990} that for radially symmetric topography located far from the axis of rotation, equation (\ref{eq:lub}) may be simplified to
\begin{equation}
\label{eq:7}
h^3(h'''+s''')+ (h^3-1)/\lambda^3_{flow} = 0.
\end{equation}
This ordinary differential equation was solved numerically on the interval $x\in[-L,L]$ using the scripting language \textsc{Matlab}.  The equation for $h$ was recast as a set of three first-order differential equations subject to the boundary conditions $h(-L)=h(L)=1$, $h'(L)=0$, and then supplied to the in-built function \texttt{bvp5c} for solution.  The sharp step occurring at the edge of the wire was approximated by \cite{Stillwagon1990} $\tfrac{\pi}{2}+\mathrm{atan}(x/(aw)) $, with $a=0.005$.  The third-order derivative $s'''$ was obtained from this function analytically.  For small values of $w$, the accurate numerical solution of equation (\ref{eq:7}) becomes increasingly time consuming.  Thus, following ref. \onlinecite{Stillwagon1990}, equation (\ref{eq:7}) was reformulated on the variable $\delta = z/h_f-h_f$ and then expanded to first order to yield $\delta''' +3\delta/\lambda^{3}_{flow}=0$.  This equation was then solved analytically for the case of an ideal step ($a\rightarrow 0 $).  The result of the approximate analytical solution was compared with the numerical solution for $1\u{\mu m}<w<10\u{\mu m}$, and found to agree to within 5\% over this range of $w$.  The data in figures \ref{fig:cartoon}b and \ref{fig:cartoon}c relating to values of $w<1\u{\mu m}$ was computed from the analytical solution.

\subsection{Spin Coating}
Substrates with topography were prepared from silicon wafers using optical lithography.  Following exposure and development of the resist, the pattern was transferred into silicon by etching with hydroflouric acid.  Additional substrates were  prepared by dispersing InAs nanowires onto a flat silicon surface.  Films were spin coated using a \textsc{Laurell Technologies} WS-650 spin coater.  The 950\pmma A resist solution used in the spin-coating process was prepared by \textsc{Micro Chem} from \pmma and anisole.  It was spun at 2000\u{rpm} for 80 seconds.  Concentrations below 2\% were obtained by further dilution of the as-purchased solution in Anisole.  AFM measurements were performed using a \textsc{Digital Instruments} AFM operated in tapping mode.

\subsection{tSPL patterning}

The thermal Scanning Probe Lithography (tSPL) was performed on our home-made system, which has been described in detail elsewhere \cite{Pires2010,Paul2011}.  Our homemade tSPL tool can locally remove the resist, polyphthaladehyde, by means of a tip that has a typical diameter of 10\u{nm} which is heated several hundred degrees centrigrade.   The structured deposition of the nickel layer was achieved using a standard \pmma lift-off process.  The details of the transfer of the pattern written in the PPA layer into the \pmma layer are summarised in the supplemental information (see figure S1) and provided in detail in ref. \onlinecite{Wolf2014}.  The details of the pattern-overlay process are given in ref. \onlinecite{Rawlings2014}.

\begin{acknowledgement}
The authors gratefully acknowledge the technical assistance of U. Drechsler and S. Reidt. The authors would also like to thank P. Mensch and S. Karg for providing both the nanowires and valuable advice.  Finally, the authors are indebted to C. Bollinger for her careful proof reading of the manuscript.  The research leading to these results received funding from the European Union's Seventh Framework Program FP7/2007-2013 under Grant Agreement No. 318804 (Single Nanometer Manufacturing for beyond CMOS devices - acronym SNM)
 and through the European Research Council StG no. 307079.
\end{acknowledgement}

\begin{suppinfo}
Details of the pattern transfer method, the analysis of the organic hardmask HM8006, and the pattern registration (Figures S1-S10).
\end{suppinfo}


\bibliography{IBMbib}

\end{document}

%% file: titlecase_exclude.tex
\Addlcwords{nm aboard, about above absent according to across after against ahead of along alongside amid amidst among around as as far as as well as at atop before behind below beneath beside between by by means of despite down on to (onto) opposite out of outside outside of owing to over past plus prior to regarding round save since than through throughout till times to toward under underneath until up upon with with regards to within without due to during except far from following for from in in addition to in case of in front of in place of in spite of inside inside of instead of in to into like mid minus near near to next next to notwithstanding of off on on account of on behalf of on top of both and not only but also either or neither nor whether or as as such that scarcely when as many as no sooner than rather than and but rr nor for yet so after although as as if as long as as much as as soon as as though because before even even if even though if if only if when if then inasmuch in order that just as lest now now since now that now when once provided provided that rather than since so that supposing than that though til unless until when whenever where whereas where if wherever whether which while who whoever why}